\newcommand{\CLASSINPUTtoptextmargin}{0.76in}
\newcommand{\CLASSINPUTbottomtextmargin}{1.1in}

\documentclass[conference]{IEEEtran}
\IEEEoverridecommandlockouts
\usepackage{cite}
\usepackage{svg}
\usepackage{amsmath,amssymb,amsfonts}
\usepackage{algorithmic}
\usepackage{graphicx}
\usepackage{textcomp}
\usepackage{xcolor}
\usepackage{hyperref}
\usepackage{braket}
\usepackage{physics}
\usepackage[normalem]{ulem}

\def\BibTeX{{\rm B\kern-.05em{\sc i\kern-.025em b}\kern-.08em
    T\kern-.1667em\lower.7ex\hbox{E}\kern-.125emX}}
\begin{document}

\title{  LatentQGAN: A Hybrid QGAN with Classical Convolutional Autoencoder}

\author{
\IEEEauthorblockN{Alexis Vieloszynski\textsuperscript{\textdagger*}, Soumaya Cherkaoui\textsuperscript{\textsection}, Ola Ahmad\textsuperscript{\textparagraph}, Jean-Frédéric Laprade\textsuperscript{\textdaggerdbl}, \\ Olivier Nahman-Lévesque\textsuperscript{\textdaggerdbl}, Abdallah Aaraba\textsuperscript{\textdagger}, and Shengrui Wang\textsuperscript{\textdagger}}\\

\IEEEauthorblockA{\textsuperscript{\textdagger}Departement of Computer Science, Université de Sherbrooke, Sherbrooke, Canada}
\IEEEauthorblockA{\textsuperscript{\textsection}Department of Computer and Software Engineering, Polytechnique Montréal, Montréal, Canada}
\IEEEauthorblockA{\textsuperscript{\textparagraph}Thales Digital Solutions, Montréal, Canada}
\IEEEauthorblockA{\textsuperscript{\textdaggerdbl}Institut Quantique, Université de Sherbrooke, Sherbrooke, Canada}\\
\IEEEauthorblockA{\textsuperscript{*}Email: alexis.vieloszynski@usherbrooke.ca}
}

\newcommand{\ola}[1]{\textit{\color{blue}{[Ola: #1]}}}
\newcommand{\jflcomment}[1]{\textit{\color{magenta}{[JFL: #1]}}}
\newcommand{\olivier}[1]{\textit{\color{olive}{[ONL: #1]}}}
\newcommand{\abdallah}[1]{\textit{\color{orange}{[AA: #1]}}}

\maketitle

\begin{abstract}
Quantum machine learning consists in taking advantage of quantum computations to generate classical data.
A potential application of quantum machine learning is to harness the power of quantum computers for generating classical data, a process essential to a multitude of applications such as enriching training datasets, anomaly detection, and risk management in finance.
Given the success of Generative Adversarial Networks in classical image generation, the development of its quantum versions has been actively conducted. However, existing implementations on quantum computers often face significant challenges, such as scalability and training convergence issues.
To address these issues, we propose LatentQGAN,
a novel quantum model that uses a hybrid quantum-classical GAN coupled with an autoencoder.
Although it was initially designed for image generation, 
the LatentQGAN approach holds potential for broader application across various practical data generation tasks.
Experimental outcomes on both classical simulators and noisy intermediate scale quantum computers have demonstrated significant performance enhancements over existing quantum methods, alongside a significant reduction in quantum resources overhead.

\end{abstract}


\section{Introduction}

Generative Adversarial Networks \cite{goodfellow2020generative} (GANs) have 
become widely recognized as an effective approach for data generation \cite{smith2020conditional}, anomaly detection \cite{schlegl2017unsupervised} as well as other common applications in machine learning.
It learns in an unsupervised way to generate new data with the similar statistics to those of the training set.
A GAN consists of two neural networks, a generator and a discriminator. The purpose of the generator is to generate new (fake) data that imitates the distribution of the training dataset, while the purpose of the discriminator is to distinguish between real data and the generated fake ones. The two neural networks are trained in competition, with the generator trying to fool the discriminator and the discriminator trying to distinguish between true and fake data.

In recent theoretical studies, it has been suggested that quantum generative models could demonstrate an advantage compared to classical equivalents \cite{lloyd2018quantum,gao2018quantum,romero2021variational}.
With the improvement of quantum computers,
much work has been done by the quantum computing community towards implementing Quantum Generative Adversarial Networks (QGANs) \cite{huang2021experimental,silver2023mosaiq}.
Among the existing implementations, the ability to effectively run the training process on real quantum computers and that to represent the high-dimensional data distribution are among the main challenges, the latter being known as a scalability issue. 
Moreover, existing QGANs suffer from lack of ability to generate diverse data that adheres to the desired data distribution, which is known as mode collapse \cite{ding2022take}.

Motivated by these challenges, we propose in this paper a new model, named LatentQGAN, to address scalability and mode collapse issues in an attempt to generate high dimensional data. We focused our effort on facilitating the training and model evaluation  on real quantum computers, which resulted in a  significant improvement over previous works.
The contributions of this paper are as follows:

(i) We propose a hybrid quantum-classical model for data generation, applied on images. The new model integrates an autoencoder and maps images into a latent space with reduced dimensionality which makes the new representation more compatible with the quantum generator.

(ii) By learning a compressed representation of the original dataset, the new model maximizes the efficiency of quantum circuits while minimizing the utilization of quantum resources such as the quantum circuit depth or the number of qubits. This allows us to circumvent the limitations of noisy intermediate scale quantum (NISQ) computers such as decoherence and limited qubit connectivity, enabling training of the model on quantum computers.

(iii) Our experiments, conducted both on simulators and on real quantum machines, show that LatentQGAN outperforms the existing implemented QGANs, and classical counterparts with the same number of parameters. The model has been trained on the MNIST dataset \cite{deng2012mnist}, which is commonly used in both classical and quantum machine learning evaluations \cite{daniel2024handwritten,goodfellow2020generative}.

\section{Background}

GANs \cite{goodfellow2020generative} are characterized by a dueling interplay between two neural networks: the generator $G$ and the discriminator $D$. This dynamic is encapsulated in a minimax game scenario, where the generator strives to produce synthetic data samples $x$ from random noise $z$, aiming to minimize the likelihood of being discerned as fake by the discriminator. Conversely, the discriminator endeavors to distinguish between real and synthetic samples, seeking to maximize its classification accuracy. D has to classify the fake data $G(z)$ as 0 to indicate it is fake, and real data $x$ as 1 to indicate it is real. Depending on the output, the parameters are updated, constituting a single iteration in the training process. This adversarial process is governed by a loss function $
\mathcal{L}(D,G)$, defined as:
\begin{equation}
    \begin{aligned}
    \min_G \max_D\mathcal{L}(D, G) &=  \mathbb{E}_{x\sim p_{\text{data}}(x)}[\log D(x)] \\
    &+ \mathbb{E}_{z\sim p_z(z)}[\log(1 - D(G(z)))],
    \end{aligned}
    \label{eq:GAN-loss-fn}
\end{equation}
where $p_{data}(x)$ represents the distribution of real data and $p_z(z)$ denotes the distribution of the noise. The loss function drives the networks to optimize their respective objectives: the discriminator's loss $L_D$ and the generator's loss $L_G$. 

Autoencoders \cite{bank2023autoencoders} represent a cornerstone in the domain of unsupervised learning, particularly in the realm of image processing and computer vision. Derived from traditional autoencoders, convolutional autoencoders (CAEs) leverage convolutional layers to capture spatial hierarchies and extract meaningful features from input data\cite{guo2017deep}. At their core, CAEs consist of two main components: the encoder and the decoder. The encoder utilizes convolutional layers that reduce spatial dimensions of the input data while extracting relevant features into what is called a latent space. Conversely, the decoder employs transposed convolutional layers (also known as deconvolution or upsampling layers) to reconstruct the original input from the encoded representation.

The training objective of CAEs involves minimizing a loss function that measures the discrepancy between the input and the reconstructed output. Typically, the mean squared error (MSE) loss is employed for this purpose.
CAEs find widespread applications in various domains, including image denoising \cite{gondara2016medical}, compression \cite{al2018convolutional}  along with other widespread applications in artificial intelligence. Moreover, they serve as a fundamental building block for more advanced techniques such as image generation and anomaly detection.

Quantum computing has witnessed a surge in interest in recent years, mainly due to three factors: scientific and engineering breakthroughs in hardware design and fabrication \cite{chen2016measuring}, the availability of quantum processors through cloud access, and the development of variational algorithms which provide a framework to experiment with the limited resources of NISQ quantum devices. Variational quantum algorithms~\cite{cerezo2021variational} (VQAs) proceed by optimizing a cost function $C(\boldsymbol{\theta})$ which encodes the problem of interest. Here, $\boldsymbol{\theta}$ is a set of real parameters, and the solution is obtained by a minimization process,
\begin{equation}
\begin{aligned}
\boldsymbol{\theta}^*  ={\textrm{arg}\min_{\theta}}~C(\boldsymbol{\theta}).
\end{aligned}
\end{equation}
The optimization routine is performed on a classical computer while the evaluation of the cost function is offloaded to a quantum computer. We can express this cost function as the expectation value of some observable $\mathcal{O}$, with respect to a parameter-dependent state,
\begin{equation}
\begin{aligned}
\expval{\mathcal{O}} = \textrm{Tr}\left[ \mathcal{O} U(\boldsymbol{\theta}) \rho U^{\dagger}(\boldsymbol{\theta}) \right].
\end{aligned}
\end{equation}
Here, $\rho$ is any initial state, generally taken as the zero state of $N$ qubits, $\rho = \ket{0}^{\otimes N}\bra{0}^{\otimes N}$, and $U(\boldsymbol{\theta})$ is a parameterized unitary transformation.

Machine learning is particularly well suited to the VQA framework, giving rise to the field of quantum machine learning (QML). For tasks such as data classification~\cite{havlivcek2019supervised}, it is common to express the parameterized quantum circuit as a composition of two unitaries: a data embedding circuit $V_{x}(\boldsymbol{x}, \boldsymbol{\phi})$ which depends on a data sample $\boldsymbol{x}$ and may include tunable parameters $\boldsymbol{\phi}$, as well as an ansatz $U(\boldsymbol{\theta})$ which is trained to identify a separating hyperplane in the Hilbert space. For generative learning~\cite{benedetti2019generative}, $V(\boldsymbol{x}, \boldsymbol{\phi})$ can be used to initialize a quantum state with some random noise and $U(\boldsymbol{\theta})$ is expected to learn the data distribution model.

\section{Related Works}

LatentGAN \cite{prykhodko2019novo} is a deep learning method for generating new data structures by combining an autoencoder with a GAN. First the autoencoder is trained to understand the dataset and create a compressed representation. Then, this compressed representation is used as a target for the generator of the GAN in its training process. LatentGAN can generate data resembling those in the dataset while also creating novel structures with comparable characteristics, making it a useful tool for generating diverse and new data structures. Training models on reduced-dimensional data is a well-established technique, enhancing computational efficiency and mitigating overfitting\cite{howley2005effect}.
\begin{figure*}
\centerline{\includegraphics[width=1\textwidth]{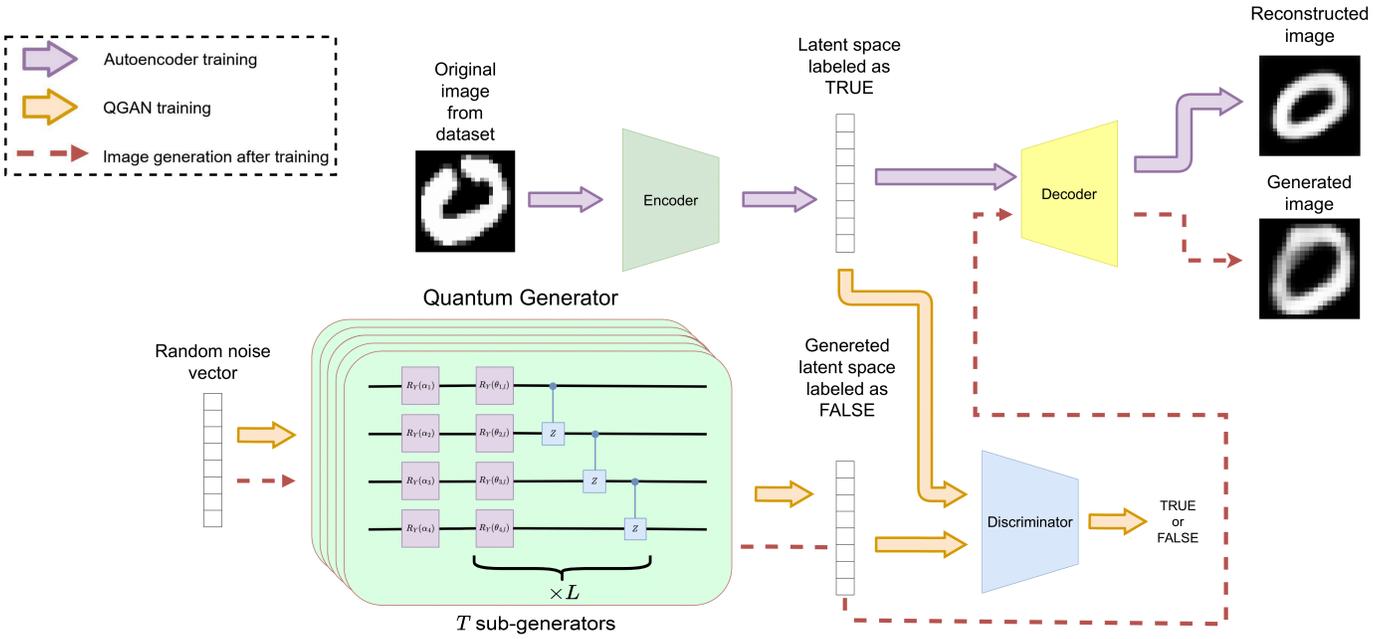}}
\caption{LatentQGAN's overall framework.}
\label{fig5}
\end{figure*}

A recent development in quantum GANs is the Quantum Patch GAN (QPatchGAN)\cite{huang2021experimental} which introduces a novel hybrid architecture where the generator is implemented with a quantum model, while the discriminator remains classical. Following an approach inspired by theoretical studies,\cite{lloyd2018quantum,gao2018quantum,romero2021variational},
QPatchGAN aims to bridge the gap between quantum advantage and the limitations of current quantum computers such as the lack of qubits, and short decoherence times. The quantum generator consists of multiple circuits, each generating a part of the image (data), with parameters updated based on the discriminator's output. Although QPatchGAN shows promising results on the $8\times8$ MNIST dataset, it struggles with scalability and mode collapse issues \cite{silver2023mosaiq}, on data with higher dimensions.

To address the aforementioned issues, a recent work, MosaiQ\cite{silver2023mosaiq} provides an alternative approach. Instead of working directly with images, MosaiQ uses Principal Components Analysis (PCA) to reduce dimensionality. Initially, a class is selected from the dataset and its dimensionality is reduced by using PCA. The reduced class representation is utilized as the target for data generation by the generator.  Furthermore,  MosaiQ makes use of adaptive noise to mitigate mode collapse, adjusting the input random vectors' range dynamically based on the generator's performance. However, current quantum computing limitations restrict MosaiQ's training to simulators, due to high number of iterations required to reach the best quality results.

During the submission process of this work, a similar architecture to ours was proposed in \cite{chang2024latent}. 
However, our approach distinguishes itself in the autoencoder implementation and advances
both the classical and quantum components of the QGAN,
offering a distinct perspective. Furthermore, we present results on real quantum hardware, demonstrating concrete evidence of the viability of this architecture on contemporary quantum devices.

\section{LatentQGAN}

\subsection{The Framework of LatentQGAN}

In this section, we describe our LatentQGAN method that offers a new approach to data generation with greater scalability compared to QPatchGAN. Moreover, LatentQGAN can be run on contemporary quantum computers. As illustrated in Fig.~\ref{fig5}, our model is composed of a convolutional autoencoder which is trained first on all classes together, and a QGAN which is trained subsequently on 
the latent representation of each class separately. To make the latent representation compatible with the quantum generator, we normalize it line by line, following the final layer of the encoder. After the completion of the training of the autoencoder, we train our hybrid QGAN by injecting random noise vectors into the generator. Once the GAN's training is finished, the generator's output is given as an input for the decoder, to create new images.
The following two sections are dedicated to describing in detail how the QGAN and the autoencoder of LatentQGAN work.

\subsection{Quantum Generative Adversarial Networks (QGAN)}
Inspired by previous work\cite{huang2021experimental}, the QGAN in LatentQGAN is hybrid with the generator being a series of quantum circuits, and the discriminator being a classical fully connected neural network.

\subsubsection{Quantum Generator}
The quantum generator comprises a set containing a number
$T$ of quantum circuits, each designed to generate and reproduce its respective portion of the data, which is divided into $T$ parts. The data is split into $T$ parts, and each generator is used to reproduce a part of the data. Each generator is a parametrized quantum circuit (PQC) as in the quantum generator part of Fig.\ref{fig5}, while each circuit is composed of a number of qubits, denoted by $N$, with an input layer of rotation gates that encode the random noise. The output is denoted by the state $\ket z$, where $\ket z =\bigotimes^N_{i=1}R_Y(\alpha_i)$ and $\alpha$ is the vector of random noise sampled uniformly.
We employ $L$ parametrized layers, which are composed of rotation gates on each qubit, with controlled-Z gate that are sequentially applied between each pair of consecutive qubits, such that each qubit controls the next qubit $U_l=\bigotimes^N_{i=1}R_Y(\theta_i)CZ_s$, 
with $CZ_s =\bigotimes^{N-1}_{i=1}CZ(i,i+1)$ and $\theta$ being the parameters to optimize.

The resulting state of each PQC can be described as $\ket{\psi_t(z)}= \left(\prod^L_{l=1}U_{l, t}\right)\ket z$ with $t \in \left[1,T\right]$.
In our model we use $N_A$ ancillary qubits to perform a non-linear transformation. The non-linearity comes from the fact that we perform a partial measurement $\prod_A$ on the ancillary subsystem. Indeed, the state of each generator $G_t$ before partial measurement can be written as $\ket{\psi_t(z)} = \sum_{i=0}^{2^N-1}\beta_i\ket i$ with $\sum_{i=0}^{2^N-1}|\beta_i|^2 = 1$. The resulting generator state after the partial measurement is then:
\begin{equation}
\rho_t(z) = \sum_{i,l=0}^{2^{N_G}-1}\frac{\beta_{i\times2^{N_A}}\times\beta_{l\times2^{N_A}}^*}{\sum_{k=0}^{2^{N_G}-1}|\beta_{k\times2^{N_A}}|^2}\ket i\bra l.
\label{eq:partial-measurement}
\end{equation} 
Each generated component is obtained by measuring each state of $\rho_t(z)$ in the computation basis $\left[\ket j\right]_{j=0}^{2^{N_G-1}}$ with j representing the j-th component value:
\begin{equation}
P_t(J = j) = Tr(\ket j \bra j \rho_t(z)) = \frac{|\beta_{j\times2^{N_A}}|^2}{\sum_{k=0}^{2^{N_G}-1}|\beta_{k\times2^{N_A}}|^2}.
\label{eq:final feature}
\end{equation} 
The output generated by the sub-generator $G_t$ is obtained by gathering each generated component: 
\begin{equation}
G_t(z) = \left[P_t(J = j) \right]_{j=0}^{2^{N_G}-1}.
\label{eq:generated output}
\end{equation} 
The final output for the generator $G(z)$ is:
\begin{equation}
G = \left[G_t(z) \right]_{i=1}^T.
\label{eq:final output}
\end{equation} 
Such an operation cannot be done on a quantum computer since it does not allow such partial measurement. So to get $P_t(J = j)$, we measure  $\ket{\psi_t(z)}$ in the complete set of computation bases and perform a post-selection by selecting the results where the ancilary qubits equal $\ket 0^{\otimes{N_A}}$. Then we perform the normalization. 

\subsubsection{Classical Discriminator} The discriminator is built using fully connected neural networks (FCNN) by taking either the latent representation of true training data $x$ or fake generated data $G(z)$ as input. Sigmoid activation ensures the output is a scalar value between 0 and 1, indicating confidence in the input's trueness, with ReLU mapping for non-linearity.

\begin{figure}

\centerline{\includegraphics[width=1\columnwidth]{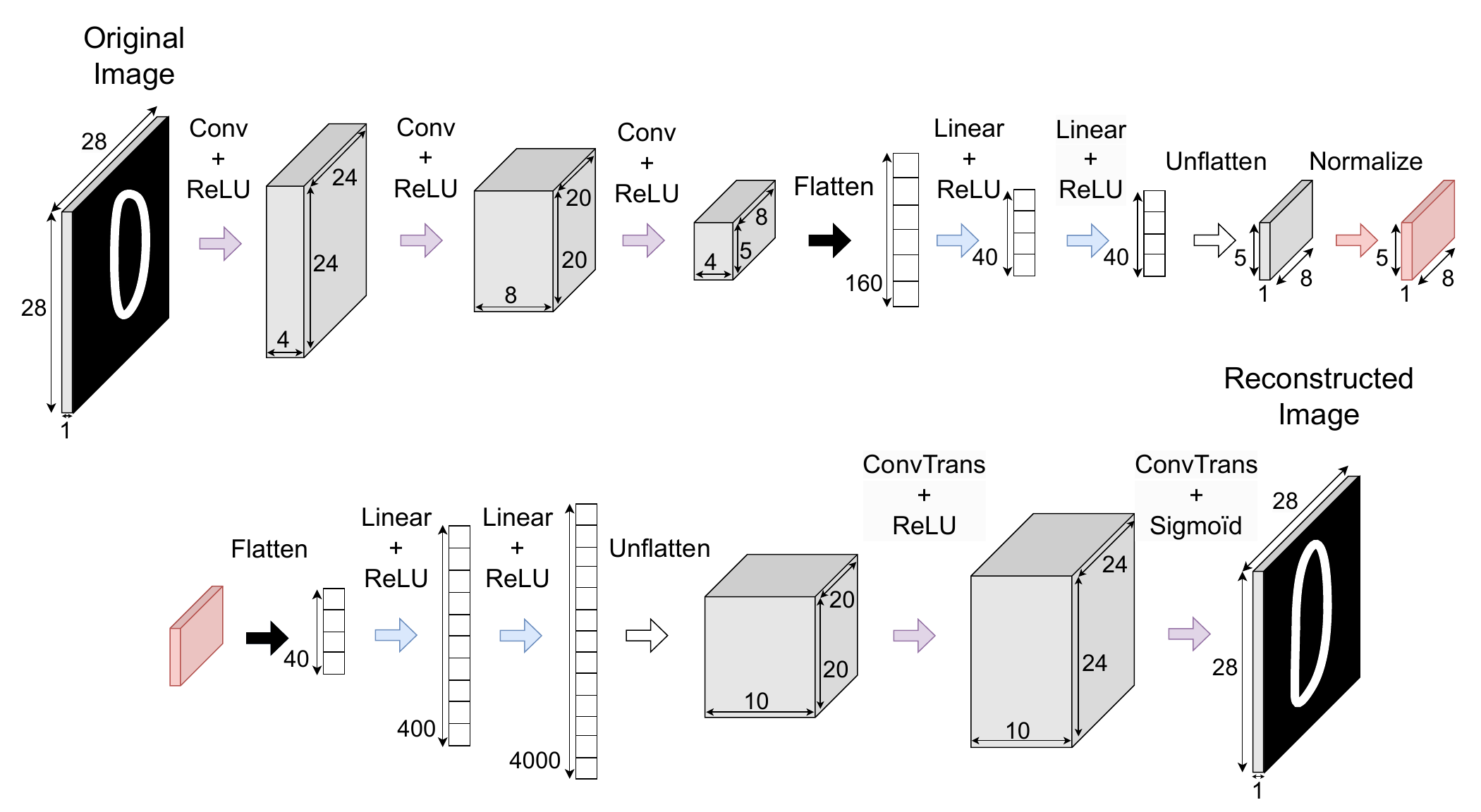}}
\caption{The autoencoder used in LatentQGAN, with the encoder on the top part, and the decoder on the bottom. The "Normalize" process is represented in red at the end of the encoder.}
\label{fig6}
\end{figure}

\subsection{The Autoencoder}

As mentioned before, for each sub-generator $t \in [1,T]$ output, there is a constraint formulated as follows:
\begin{equation}
\sum_{j=0}^{2^{N_G}-1}P_t(J = j) = 1 
\label{eq:generator constraint}
\end{equation} 

 We have to make the latent representation of the encoder compatible with the generator's set of constraints. We use a normalization per line on the latent representation, named "Normalize" on Fig.~\ref{fig6}, which is formulated as follows: 
let $h$ and $\hat h$ be the latent representation respectively before and after the normalize process. We represent $h$ by the matrix $h = (h_{i,j})$ where $i \in [1,T]$ and $j \in [1,2^{N_G}]$. After the "Normalize" process, we can represent $\hat h$ by:
\begin{equation}
\hat h = \left(\frac{h_{i,j}}{\sum_{j=1}^{2^{N_G}}h_{i,j}}\right),
\label{eq:normalized LR}
\end{equation}
which satisfies $\sum_{j=1}^{2^{N_G}} \hat h_{i,j} = 1$.

In the decoder, the initial linear layers take the latent representation produced by the encoder and begin to reconstruct it into a form compatible with the original data. The transpose convolutional layers reverse the encoding process, gradually expanding the latent representation to reconstruct the original data in its original dimensions. This demonstrates another contribution of using an autoencoder which is that it is able to turn constrained output of the quantum generator into generalized data. ReLU activation function, while the sigmoid activation function in the final layer ensures that the reconstructed values fall within the $\left[0, 1\right]$ range, consistent with the scale of the original data.
By combining the encoder and decoder, the model aims to learn a compressed representation of the input data in the latent space while being able to faithfully reconstruct the original data from this representation. 

\begin{figure*}[ht!] 
\centerline{\includegraphics[width=0.9\textwidth ]{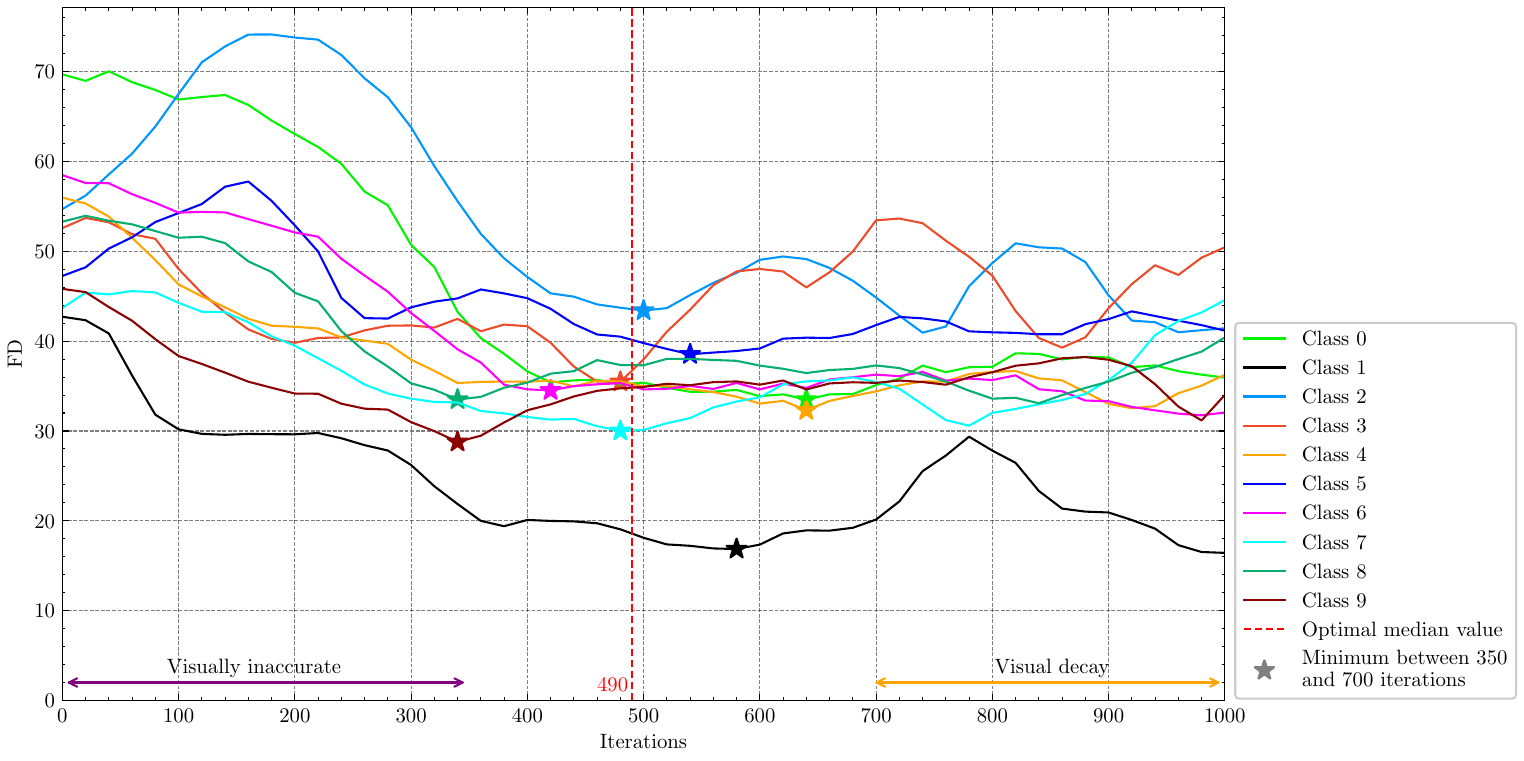}}

\caption{FD results over training.
The Fig.\ref{fig9} shows a visual inaccuracy and decay respectively before 350 iterations and after 700 iterations}

\label{fig8}
\end{figure*}

\section{Experimental details and results}

\subsection{Experimental details}

\begin{figure}[ht!]
\centerline{\includegraphics[width=0.8\columnwidth, keepaspectratio]{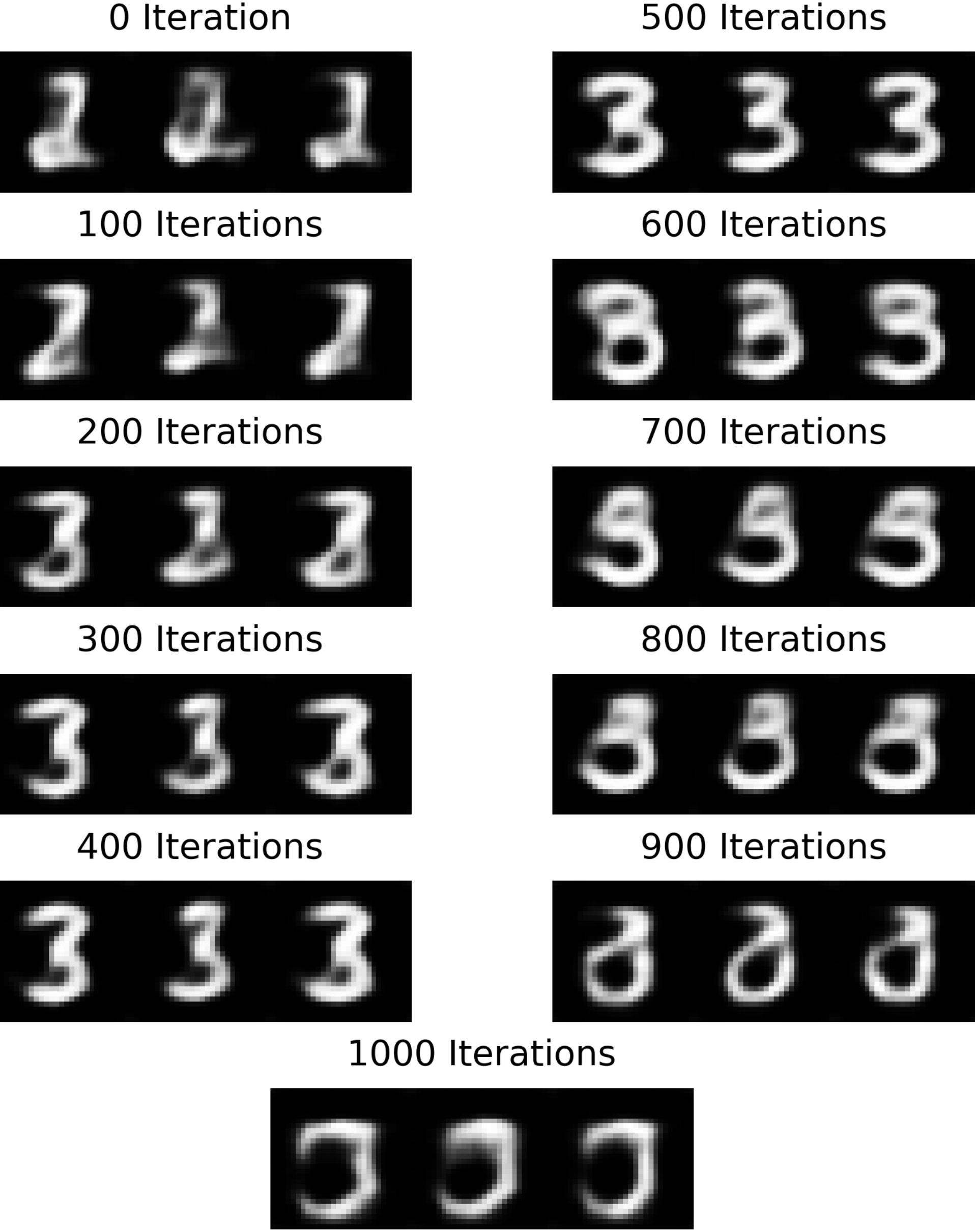}}
\caption{Images generated by LatentQGAN for class 3 over 1000 training iterations, on simulator. Before 300 iterations, the results are visually inaccurate, and after 700 iterations, the results visually decay.}
\label{fig9}
\end{figure}

We benchmark LatentQGAN on the MNIST dataset, which consists of $28 \times 28$ gray scale images depicting handwritten digits ranging from 0 to 9. The encoder compresses the information contained in the images into a normalized latent space of dimension $5\times8$. To match this dimension, the quantum generator is composed of five quantum circuits, one for each row of the latent space, with $N_G = 3$ and $N_A = 1$ for the generator and ancillary qubits respectively. In our experiments, the variational circuits we used were composed of $L=7$ layers, for a total of 140 parameters. These values are the result of a careful search in hyperparameter space and represent an optimal trade-off between maximizing performance of the model and minimizing computational time on the quantum machine.
Employing five circuits of four qubits each to form an array of size eight is computationally efficient, as they can be parallelized on a single 127-qubit IBM eagle quantum processor, making this model trainable on contemporary quantum computers,

\begin{figure*} 
\centerline{\includegraphics[width=1\textwidth, keepaspectratio]{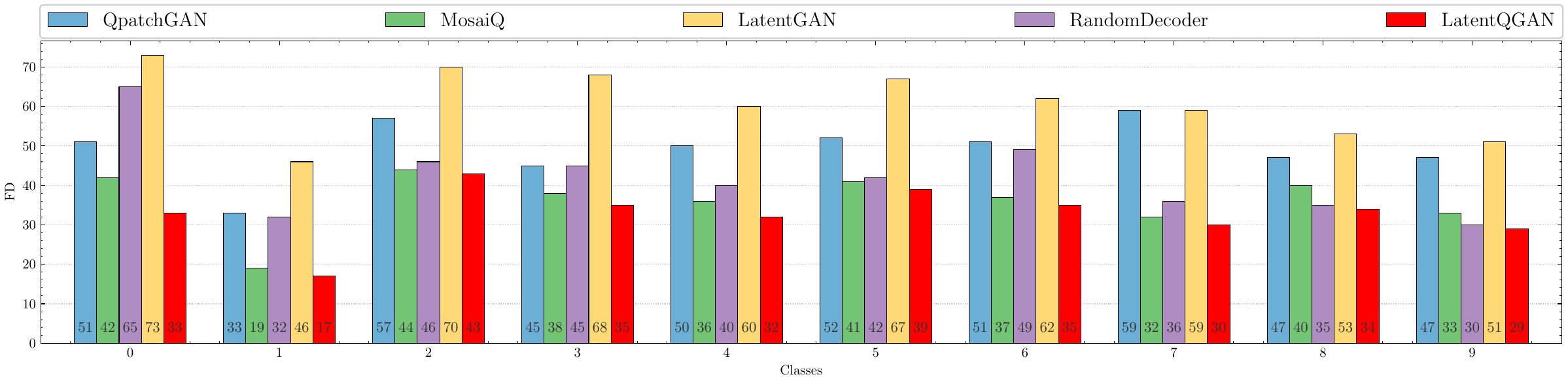}}
\caption{FD evaluated on many models for comparison with LatentQGAN, on simulator}
\label{fig7}
\end{figure*}

On the other hand, the discriminator is built using a neural network with FCNNs, ReLU activation after convolutional layers and a sigmoid function for the output layer. The FCNN has an input layer of 40 neurons, followed by two hidden layers with 64 and 16 neurons, and an output layer with 1 neuron, totalling 3681 parameters. 
To compute the gradients of the generator, we employ the parameter-shift rule\cite{schuld2019evaluating}.
LatentQGAN uses the stochastic gradient descent optimizer and the binary
cross entropy loss, described in \eqref{eq:GAN-loss-fn} for the shared loss of the generator and discriminator and the mean squared error loss for the autoencoder.
The autoencoder's learning rate is 0.05, with a batch size of 20 and its training is done on 100 epochs, which we have determined empirically and led to optimal results. The generator learning rate is 0.3 and the discriminator learning rate is 0.01 as recommended in \cite{silver2023mosaiq,huang2021experimental} with a batch size of 1 to limit the training time. The training is done on a number of iteration that depends on the class, based on our experiments, the optimal number of iterations for all 10 MNIST classes is 490.

In this study, model evaluation was performed based on the Fréchet Distance (FD) metric, which is commonly used to evaluate QGANs \cite{huang2021experimental,silver2023mosaiq}. The FD metric serves as a measure of the similarity between two distributions of data by computing the distance between their corresponding multivariate Gaussian distributions. Mathematically, FD is calculated as follows:
\begin{equation}
    FD=\lVert\mu_r - \mu_g \rVert_2^2+Tr(\Sigma_r+\Sigma_g - 2(\Sigma_r\Sigma_g)^{\frac{1}{2}}),
\end{equation}
where $\mu_r$ and $\mu_g$ are the mean vectors, and $\Sigma_r$ and $\Sigma_g$ are the covariance matrices of the real and generated data distributions, respectively. A lower FD score indicates a higher resemblance between the generated samples and the real data distribution. While useful for gauging the similarity between generated and real data distributions, the FD does not always correspond to human evaluations as shown in Fig.\ref{fig10}. Thus, though valuable, it should be supplemented with further evaluations, including perceptual studies where human observers rate the quality of generated images.

To evaluate our model, we conducted comparisons with QPatchGAN and MosaiQ, both previously introduced. Additionally, we compared our model to LatentGAN, the fully classical version of LatentQGAN with an equivalent number of parameters in each component, including the generator. Finally, we compared all the approaches to a model called RandomDecoder, which is a model that generates images by feeding random normalized noise vectors into the decoder. These comparisons allow demonstrating the better results of our model and the importance of learning the latent space for achieving satisfactory results.

\subsection{Experimental results}
\begin{figure} 
\centerline{\includegraphics[width=\columnwidth, keepaspectratio]{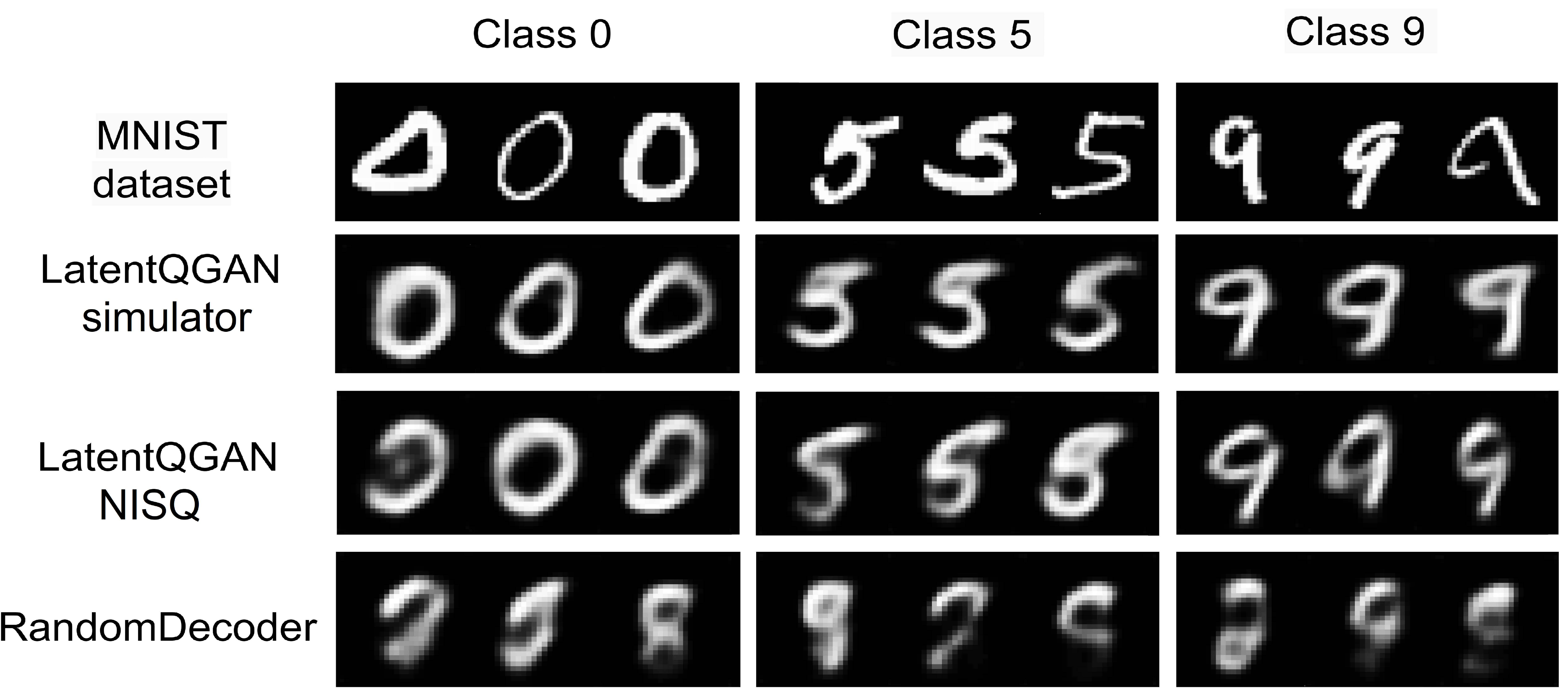}}
\caption{Generated images with the lowest FD for LatentQGAN on the simulator.
Images generated on NISQ needed 550 iterations for class 9 and 5, and 650 for class 0 to achieve minimal FD, out of 800 iterations.}
\label{fig10}
\end{figure}

Model evaluation involved a comparative analysis between generated samples and ground truth data. As we can see in Fig.~\ref{fig8}, based on our experiments, the minimum FD was obtained after 350 to 700 iterations. As depicted in Fig.\ref{fig9}, prior to 350 iterations, the model fails to generate visually accurate images. Subsequently, between 350 and 700 iterations, the generated images improve noticeably. However, beyond 700 iterations, there is a gradual decline in visual quality.

In the results shown in Fig.~\ref{fig7} and Fig.~\ref{fig10}, LatentQGAN consistently outperforms other models like QPatchGAN and MosaiQ in terms of image quality, both visually and quantitatively, in simulations. 
Interestingly, LatentQGAN achieves this advantageous comparison with only five 4-qubits quantum circuits and a total of 140 parameters, while MosaiQ uses eight 5-qubits circuits and 240 parameters.
Additionally, LatentQGAN requires 676 times fewer iterations than MosaiQ to achieve comparable results. Moreover, although the FD of RandomDecoder remains low, the visual outcomes shown in Fig.\ref{fig10} suggest an inadequate representation of the original distribution. 

\begin{table}
\caption{Comparison of LatentQGAN's FD on Quantum Simulator vs. Quantum Computer. Results on the quantum computer are obtained from a single run using identical noise vectors and initial parameters as those used in simulator trainings.}
\begin{center}
\begin{tabular}{|c|c|c|c|}
\hline
\textbf{} & \textbf{\textit{Class 0}}& \textbf{\textit{Class 5}}& \textbf{\textit{Class 9}} \\
\hline
Quantum simulator&33&39&29  \\
\hline
Quantum computer&38&43&34  \\
\cline{1-4} 

\end{tabular}

\label{tab1}

\end{center}

\end{table}

Experimentally, similar models\cite{chang2024latent} tend to outperform other QGANs in generating high dimensional images from 3 data sets (MNIST, fashion MNIST and SAT4) on simulator. However, this accuracy requires significantly more parameters—ranging from 10 to 70 times more than our quantum generator—rendering it untrainable on current NISQ devices within a reasonable computation time.

Our experiments were conducted on \textit{ibm-quebec} between April $1^{st}$ and May $1^{st}$ 2024, using 2048 shots and M3 error mitigation \cite{nation2021scalable}. Given the long wait times and limited availability of quantum computers, we are focusing on presenting results for the classes $0$, $5$ and $9$. These classes are chosen to represent a wide range of complexities in terms of shape and information. The results obtained on the quantum computer, as depicted in Table.\ref{tab1} and Fig.~\ref{fig10} , align with expectations: they exhibit slightly reduced accuracy compared to simulations due to noise. Despite this, the results of LatentQGAN on NISQ are visually recognizable and competitive with simulation results. 
While a significant amount of learning is done by the autoencoder, Fig. \ref{fig8} and \ref{fig9} demonstrate that the generator often requires multiple iterations before the generated images reach acceptable visual and quantitative quality. Moreover, images generated by RandomDecoder, as shown previously, are less representative of the original distribution than those of LatentQGAN, suggesting the model’s inability to generate visually accurate results without a proper learning of the patterns in the latent representation.
Additionally, the original LatentGAN with the same number of parameters fails to reproduce the results of our model, regardless of whether there is normalization in the encoder or not. This discrepancy suggests that factors beyond the autoencoder contribute significantly to the model's performance.

\section{Conclusion}

 The LatentQGAN proposed in this paper allows the QGAN model to learn more effectively a latent representation of data generated by an encoder from an autoencoder. This method enables dealing with high-dimensional data and training of the QGAN on real quantum computers, and demonstrates significant improvement in the quality of data generation both on quantum simulators and real quantum computers. 
Using autoencoders to reduce input data dimensionality could enable more effective use of quantum machine learning models. Exploring this approach could prove valuable, as it might enhance the versatility of quantum systems in tackling diverse tasks.
 Extending the model to a full quantum implementation is worth exploring. This work will also be utilized as a basis for our further investigations into generation of complex data types, in particular time series data generation and anomaly detection, where a major challenge will be the non-stationarity of data.

\section*{Acknowledgments}
This work was supported by the Ministère de l’Économie et de l’Innovation du Québec through its contribution to the Quantum AlgoLab of Institut quantique at Université de Sherbrooke. The authors would also like to extend their gratitude to the Natural Sciences and Engineering Research Council of Canada, Prompt Innov, Thales Digital Solutions, and Zetane Systems for their financial support of this research.

\bibliographystyle{IEEEtran}

\bibliography{references}

\begin{thebibliography}{10}
\providecommand{\url}[1]{#1}
\csname url@samestyle\endcsname
\providecommand{\newblock}{\relax}
\providecommand{\bibinfo}[2]{#2}
\providecommand{\BIBentrySTDinterwordspacing}{\spaceskip=0pt\relax}
\providecommand{\BIBentryALTinterwordstretchfactor}{4}
\providecommand{\BIBentryALTinterwordspacing}{\spaceskip=\fontdimen2\font plus
\BIBentryALTinterwordstretchfactor\fontdimen3\font minus \fontdimen4\font\relax}
\providecommand{\BIBforeignlanguage}[2]{{%
\expandafter\ifx\csname l@#1\endcsname\relax
\typeout{** WARNING: IEEEtran.bst: No hyphenation pattern has been}%
\typeout{** loaded for the language `#1'. Using the pattern for}%
\typeout{** the default language instead.}%
\else
\language=\csname l@#1\endcsname
\fi
#2}}
\providecommand{\BIBdecl}{\relax}
\BIBdecl

\bibitem{goodfellow2020generative}
I.~Goodfellow, J.~Pouget-Abadie, M.~Mirza, B.~Xu, D.~Warde-Farley, S.~Ozair, A.~Courville, and Y.~Bengio, ``Generative adversarial networks,'' \emph{Communications of the ACM}, vol.~63, no.~11, pp. 139--144, 2020.

\bibitem{smith2020conditional}
K.~E. Smith and A.~O. Smith, ``Conditional gan for timeseries generation,'' \emph{arXiv preprint arXiv:2006.16477}, 2020.

\bibitem{schlegl2017unsupervised}
T.~Schlegl, P.~Seeb{\"o}ck, S.~M. Waldstein, U.~Schmidt-Erfurth, and G.~Langs, ``Unsupervised anomaly detection with generative adversarial networks to guide marker discovery,'' in \emph{International conference on information processing in medical imaging}.\hskip 1em plus 0.5em minus 0.4em\relax Springer, 2017, pp. 146--157.

\bibitem{lloyd2018quantum}
S.~Lloyd and C.~Weedbrook, ``Quantum generative adversarial learning,'' \emph{Physical review letters}, vol. 121, no.~4, p. 040502, 2018.

\bibitem{gao2018quantum}
X.~Gao, Z.-Y. Zhang, and L.-M. Duan, ``A quantum machine learning algorithm based on generative models,'' \emph{Science advances}, vol.~4, no.~12, p. eaat9004, 2018.

\bibitem{romero2021variational}
J.~Romero and A.~Aspuru-Guzik, ``Variational quantum generators: Generative adversarial quantum machine learning for continuous distributions,'' \emph{Advanced Quantum Technologies}, vol.~4, no.~1, p. 2000003, 2021.

\bibitem{huang2021experimental}
H.-L. Huang, Y.~Du, M.~Gong, Y.~Zhao, Y.~Wu, C.~Wang, S.~Li, F.~Liang, J.~Lin, Y.~Xu, R.~Yang, T.~Liu, M.-H. Hsieh, H.~Deng, H.~Rong, C.-Z. Peng, C.-Y. Lu, Y.-A. Chen, D.~Tao, X.~Zhu, and J.-W. Pan, ``Experimental quantum generative adversarial networks for image generation,'' \emph{Physical Review Applied}, vol.~16, no.~2, p. 024051, 2021.

\bibitem{silver2023mosaiq}
D.~Silver, T.~Patel, W.~Cutler, A.~Ranjan, H.~Gandhi, and D.~Tiwari, ``Mosaiq: Quantum generative adversarial networks for image generation on nisq computers,'' in \emph{Proceedings of the IEEE/CVF International Conference on Computer Vision}, 2023, pp. 7030--7039.

\bibitem{ding2022take}
Z.~Ding, S.~Jiang, and J.~Zhao, ``Take a close look at mode collapse and vanishing gradient in gan,'' in \emph{2022 IEEE 2nd International Conference on Electronic Technology, Communication and Information (ICETCI)}.\hskip 1em plus 0.5em minus 0.4em\relax IEEE, 2022, pp. 597--602.

\bibitem{deng2012mnist}
L.~Deng, ``The mnist database of handwritten digit images for machine learning research [best of the web],'' \emph{IEEE signal processing magazine}, vol.~29, no.~6, pp. 141--142, 2012.

\bibitem{daniel2024handwritten}
R.~Daniel, B.~Prasad, P.~K. Pasam, D.~Sudarsa, A.~Sudhakar, and B.~V. Rajanna, ``Handwritten digit recognition using quantum convolution neural network,'' \emph{Int J Artif Intell}, vol.~13, no.~1, pp. 533--541, 2024.

\bibitem{bank2023autoencoders}
D.~Bank, N.~Koenigstein, and R.~Giryes, ``Autoencoders,'' \emph{Machine learning for data science handbook: data mining and knowledge discovery handbook}, pp. 353--374, 2023.

\bibitem{guo2017deep}
X.~Guo, X.~Liu, E.~Zhu, and J.~Yin, ``Deep clustering with convolutional autoencoders,'' in \emph{Neural Information Processing: 24th International Conference, ICONIP 2017, Guangzhou, China, November 14-18, 2017, Proceedings, Part II 24}.\hskip 1em plus 0.5em minus 0.4em\relax Springer, 2017, pp. 373--382.

\bibitem{gondara2016medical}
L.~Gondara, ``Medical image denoising using convolutional denoising autoencoders,'' in \emph{2016 IEEE 16th international conference on data mining workshops (ICDMW)}.\hskip 1em plus 0.5em minus 0.4em\relax IEEE, 2016, pp. 241--246.

\bibitem{al2018convolutional}
A.~Z. Al-Marridi, A.~Mohamed, and A.~Erbad, ``Convolutional autoencoder approach for eeg compression and reconstruction in m-health systems,'' in \emph{2018 14th international wireless communications \& mobile computing conference (IWCMC)}.\hskip 1em plus 0.5em minus 0.4em\relax IEEE, 2018, pp. 370--375.

\bibitem{chen2016measuring}
Z.~Chen, J.~Kelly, C.~Quintana, R.~Barends, B.~Campbell, Y.~Chen, B.~Chiaro, A.~Dunsworth, A.~Fowler, E.~Lucero, E.~Jeffrey, A.~Megrant, J.~Mutus, M.~Neeley, C.~Neill, P.~J.~J. O'Malley, P.~Roushan, D.~Sank, A.~Vainsencher, J.~Wenner, T.~C. White, A.~N. Korotkov, and J.~M. Martinis, ``Measuring and suppressing quantum state leakage in a superconducting qubit,'' \emph{Physical review letters}, vol. 116, no.~2, p. 020501, 2016.

\bibitem{cerezo2021variational}
M.~Cerezo, A.~Arrasmith, R.~Babbush, S.~C. Benjamin, S.~Endo, K.~Fujii, J.~R. McClean, K.~Mitarai, X.~Yuan, L.~Cincio, and P.~J. Coles, ``Variational quantum algorithms,'' \emph{Nature Reviews Physics}, vol.~3, no.~9, pp. 625--644, 2021.

\bibitem{havlivcek2019supervised}
V.~Havl{\'\i}{\v{c}}ek, A.~D. C{\'o}rcoles, K.~Temme, A.~W. Harrow, A.~Kandala, J.~M. Chow, and J.~M. Gambetta, ``Supervised learning with quantum-enhanced feature spaces,'' \emph{Nature}, vol. 567, no. 7747, pp. 209--212, 2019.

\bibitem{benedetti2019generative}
M.~Benedetti, D.~Garcia-Pintos, O.~Perdomo, V.~Leyton-Ortega, Y.~Nam, and A.~Perdomo-Ortiz, ``A generative modeling approach for benchmarking and training shallow quantum circuits,'' \emph{npj Quantum Information}, vol.~5, no.~1, p.~45, 2019.

\bibitem{prykhodko2019novo}
O.~Prykhodko, S.~V. Johansson, P.-C. Kotsias, J.~Ar{\'u}s-Pous, E.~J. Bjerrum, O.~Engkvist, and H.~Chen, ``A de novo molecular generation method using latent vector based generative adversarial network,'' \emph{Journal of Cheminformatics}, vol.~11, pp. 1--13, 2019.

\bibitem{howley2005effect}
T.~Howley, M.~G. Madden, M.-L. O’Connell, and A.~G. Ryder, ``The effect of principal component analysis on machine learning accuracy with high dimensional spectral data,'' in \emph{International Conference on Innovative Techniques and Applications of Artificial Intelligence}.\hskip 1em plus 0.5em minus 0.4em\relax Springer, 2005, pp. 209--222.

\bibitem{chang2024latent}
S.~Y. Chang, S.~Thanasilp, B.~L. Saux, S.~Vallecorsa, and M.~Grossi, ``Latent style-based quantum gan for high-quality image generation,'' \emph{arXiv preprint arXiv:2406.02668}, 2024.

\bibitem{schuld2019evaluating}
M.~Schuld, V.~Bergholm, C.~Gogolin, J.~Izaac, and N.~Killoran, ``Evaluating analytic gradients on quantum hardware,'' \emph{Physical Review A}, vol.~99, no.~3, p. 032331, 2019.

\bibitem{nation2021scalable}
P.~D. Nation, H.~Kang, N.~Sundaresan, and J.~M. Gambetta, ``Scalable mitigation of measurement errors on quantum computers,'' \emph{PRX Quantum}, vol.~2, no.~4, p. 040326, 2021.

\end{thebibliography}

\end{document}